\documentclass[11pt]{article}
\usepackage{graphicx}
\begin{document}
%

\begin{center}
{\large \bf Leading Questions in an Extended Standard Model}

\vskip.5cm

W-Y. Pauchy Hwang$^{1,}$\footnote{Correspondence Author;
 Email: wyhwang@phys.ntu.edu.tw; arXiv:1212.4944v2 [hep-ph]
 12 April 2013; to be published in "The Universe".}
 and Tung-Mow Yan$^2$ \\
{\em $^1$Asia Pacific Organization/Center for Cosmology and Particle Astrophysics, \\
Institute of Astrophysics, Center for Theoretical Sciences,\\
and Department of Physics, National Taiwan University,
     Taipei 106, Taiwan\\
$^2$Department of Physics, Cornell University, Ithaca, N.Y. 14850}
\vskip.2cm


{\small(December 21, 2012)}
\end{center}

\begin{abstract}
We would like to discuss the language to write an extended Standard Model
- using renormalizable quantum field theory as the framework; to start
with certain basic units together with a certain gauge group. Specifically
we use the left-handed and right-handed spinors to form the basic units
together with $SU_c(3) \times SU_L(2) \times U(1) \times SU_f(3)$ as the gauge
group. We could write down the extended Standard Model, though the details
of the Higgs mechanism remains to be worked out. The same general quest
appeared about forty years ago - the so-called "How to build up a model".
It is timely to address the same question again especially since we could
now put together "Dirac similarity principle" and "Higgs minimum
hypothesis" as two additional working rules.

\bigskip

{\parindent=0pt PACS Indices: 12.60.-i (Models beyond the standard
model); 98.80.Bp (Origin and formation of the Universe); 12.10.-g
(Unified field theories and models).}
\end{abstract}

\bigskip

\section{Introduction}

As time goes by, our confidence in what we are doing seems to be
dwindling - so to look for "superstring", etc., as alternatives.
Nevertheless, the language which was built up during the last
century, namely, renormalizable quantum field theory, may still
be the language underlying the final Standard Model. It will be
the language of this paper.

Usually in a textbook, the QCD chapter precedes the one on
Glashow-Weinberg-Salam (GWS) electroweak theory. Nothing is wrong
with it but the basic units (or the building blocks) are further
divided into the left-handed and right-handed components. It would
be nice (in helping us in thinking) if the framework is formulated
all at once - in an extended Standard Model we could see everything
consistent with one another. Then, the questions which we pose
could have broader meanings and implications. Thus, this is what
we wish to do.

We shall work with the Lie group $SU_c(3) \times SU_L(2) \times
U(1) \times SU_f(3)$ as the gauge group. Thus, the basic units
are made up from quarks (of six flavors, of three colors, and of
the two helicities) and leptons (of three generations and of the
two helicities), together with all originally massless gauge
bosons and the somewhat hidden induced Higgs fields. In view of
the search over the last forty years, we could assume "minimum
Higgs hypothesis" as the working rule.

If we look at the basic units as compared to the original
particle, i.e. the electron, the starting basic units are all
"point-like" Dirac particles. Dirac invented Dirac electrons
eighty years ago and surprisingly enough these "point-like" Dirac
particles are the basic units of the Standard Model. Thus, we call
it "Dirac Similarity Principle" - a salute to Dirac; a triumph to
mathematics. Our world could indeed be described by the proper
mathematics. The proper mathematics may be the renormalizable
quantum field theory, although our confidence in it sort of
fluctuates in time.

There is no way to "prove" the above two working rules - "Dirac
Similarity Principle" and "minimum Higgs hypothesis". It might
be associated with the peculiar property of our Lorentz-invariant
space-time. To use these two working rules, we could simplify
tremendously the searches for the new extended Standard Models.

\medskip

\section{The Statement for the Extended Standard Model}

So far, we have decided on the basic units - those left-handed and
right-handed quarks and leptons; the gauge group is chosen to be
$SU_c(3) \times SU_L(2) \times U(1) \times SU_f(3)$.

In the gauge sector, the lagrangian is fixed if the gauge group
is given; only for a massive gauge theory, Higgs fields are called
for and we postpone its discussions until we have spelled out the
fermion sector.

For the fermion sector, the story is again fixed if the so-called
"gauge-invariant derivative", i.e. $D_\mu$ in the kinetic-energy
term $-\bar \Psi \gamma_\mu D_\mu \Psi$, is given for a given
basic unit \cite{Books}.

Thus, we have, for the up-type right-handed quarks $u_R$, $c_R$,
and $t_R$,
\begin{equation}
D_\mu = \partial_\mu - i g_c {\lambda^a\over 2} G_\mu^a -
i {2\over 3} g'B_\mu,
\end{equation}
and, for the rotated down-type right-handed quarks $d'_R$, $s'_R$,
and $b'_R$,
\begin{equation}
D_\mu = \partial_\mu - i g_c {\lambda^a\over 2} G_\mu^a -
i (-{1\over 3}) g' B_\mu.
\end{equation}

On the other hand, we have, for the $SU_L(2)$ quark doublets,
\begin{equation}
D_\mu = \partial_\mu - i g_c {\lambda^a\over 2} G_\mu^a - i g
{\vec \tau\over 2}\cdot \vec A_\mu - i {1\over 6} g'B_\mu.
\end{equation}

For the lepton side, we introduce the family triplet,
$(\nu_\tau^R,\,\nu_\mu^R,\,,\nu_e^R)$ (column), under $SU_f(3)$.
Since the minimal Standard Model does not see the right-handed
neutrinos, it would be a natural way to make an extension of the
minimal Standard Model. We propose that neutrinos are only species
seeing the family gauge sector. Or, we have, for $(\nu_\tau^R,\,
\nu_\mu^R,\,\nu_e^R)$ ($\equiv \Psi_R(3,1)$),
\begin{equation}
D_\mu = \partial_\mu - i \kappa {\bar\lambda^a\over 2} F_\mu^a.
\end{equation}
and, for the left-handed $SU_f(3)$-triplet and $SU_L(2)$-doublet
$((\nu_\tau^L,\,\tau^L),\, (\nu_\mu^L,\,\mu^L),\, (\nu_e^L,\,e^L))$
(all columns) ($\equiv \Psi_L(3,2)$),
\begin{equation}
D_\mu = \partial_\mu - i \kappa {\bar\lambda^a\over 2} F_\mu^a - i g
{\vec \tau\over 2} \cdot \vec A_\mu + i {1\over 2} g' B_\mu.
\end{equation}
If the right-handed charged leptons were singlets under $SU_f(3)$,
then the mass-generation terms for charged leptons would involve
the cross terms, such as $\mu\to e$, which is not acceptable at
all. Thus, the right-handed charged leptons have to form another
triplet $\Psi_R^C(3,1)$ under $SU_f(3)$.

In other words, the quark masses are given by the Higgs mechanism in 
the minimal Standard Model while the masses of charged leptons are 
determined by $\bar\Psi_L(3,2)\Psi_R^C(3,1)\Phi(1,2) + c.c.$, only a 
universal number in the leading-order sense. To make a reasonable 
theory \cite{Family}, we have to make certain that
all gauge bosons and the residual family Higgs particles are very 
massive, i.e. $\ge$ a few TeV.

As slightly differing from the previous effort \cite{Family}, we would
like to write down the $SU_c(3) \times SU_L(2) \times U(1) \times
SU_f(3)$ Standard Model {\it all at once}. We introduce the neutrino 
mass term as follows:
\begin{equation}
i {\eta\over 2} {\bar\Psi}_L(3,2) \times \Psi_R(3,1) \cdot \Phi(3,2) + h.c.,
\end{equation}
where $\Psi(3,i)$ are the lepton multiplets just mentioned above (with
the first number for $SU_f(3)$ and the second for $SU_L(2)$). The
cross (curl) product is somewhat new \cite{Family}, referring to
the singlet combination of three triplets in $SU(3)$ - an $SU(3)$ 
operation (and not a matrix product). The Higgs field
$\Phi(3,2)$ is new in this effort, because it carries some nontrivial
$SU_L(2)$ charge.

\medskip

\section{Lepton-flavor-violating Interaction}

Neutrinos have masses, the tiny masses far below the range of the masses of
the quarks and charged leptons. Neutrinos oscillate among themselves,
giving rise to a lepton-flavor violation (LFV). There are other oscillation
stories, such as the oscillation in the $K^0-{\bar K}^0$ system, but
there is a fundamental difference here - the $K^0-{\bar K}^0$ system is
composite while neutrinos are "point-like" Dirac particles. It is true that
neutrino masses and neutrino oscillations may be regarded as one of
the most important experimental facts over the last thirty years \cite{PDG}.

In fact, certain LFV processes such as $\mu \to e + \gamma$ \cite{PDG}
and $\mu + A \to A* + e$ are closely related to the most cited picture of
neutrino oscillations so far \cite{PDG}. In recent publications by
one \cite{Hwang10} of us, it was pointed
out that the cross-generation or off-diagonal neutrino-Higgs interaction
may serve as the detailed mechanism of neutrino oscillations, with some
vacuum expectation value of the new Higgs field(s).

In the other words, the first term in the last equation [Eq. (6)] can be used
as the basis to analyze the various lepton-flavor-violating decays and reactions.

To illustrate the point further, we calculate the golden lepton-flavor-violating
decay $\mu \to e + \gamma$ as the celebrated example. We show in Figures 1(a),
1(b), and 1(c) three leading basic Feynman diagrams. Here the conversion
of $\nu_\mu$ into $\nu_e$ is marked by a cross sign and it is a term from the
off-diagonal interaction given above with the Higgs vacuum expectation value
$u_0$. Here the Higgs masses are assumed to be very large, i.e.,
greater than a few $TeV$, as in $SU_f(3)$. The only small number (coupling) is
$\eta$, consistent with the tiny masses of neutrinos.

\begin{figure}[h]
\centering
\includegraphics[width=4in]{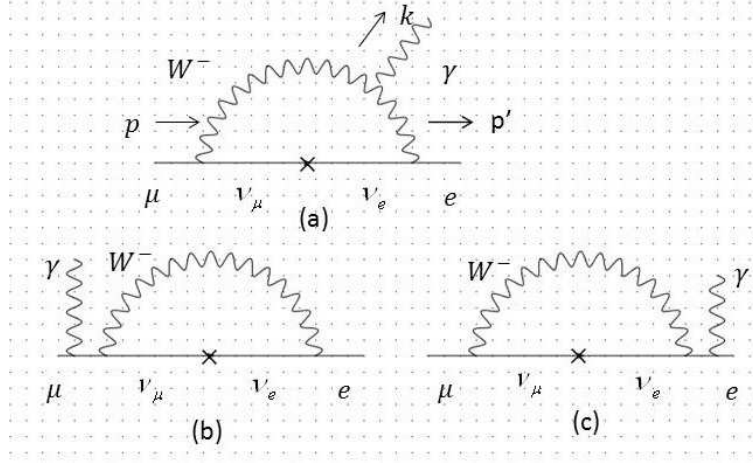}
\caption{The leading diagrams for $\mu \to e + \gamma$.}
\end{figure}

Using Feynman rules from Wu and Hwang \cite{Books}, we write, for Fig. 1(a),

\begin{eqnarray*}
{1\over (2\pi)^4} \int d^4q \cdot {\bar u}(p',s')\cdot &i \cdot {1\over 2 \sqrt 2}
{e\over sin \theta_W}\cdot i \gamma_\lambda (1+ \gamma_5)\nonumber\\
\cdot {1\over i} {m_2-i\gamma\cdot q\over {m_2^2+q^2-i\epsilon}}\cdot
&i \cdot i \eta (-)u_0 \cdot {1\over i} {m_1-i\gamma\cdot q\over
{m_1^2 + q^2-i\epsilon}} \nonumber\\
\cdot i\cdot {1\over 2 \sqrt 2}{e\over sin\theta_W}\cdot &i \gamma_{\lambda'}
(1+\gamma_5)\cdot u(p,s)\nonumber\\
\cdot {1\over i} {\delta_{\lambda'\mu}\over {M_W^2+(p-q)^2-i\epsilon}}\cdot
{\epsilon_\sigma(k)\over \sqrt{2k_0}}\cdot &\Delta_{\sigma\mu\nu} \cdot
{1\over i} {\delta_{\nu\lambda}\over {M_W^2+(p-q-k)^2-i\epsilon}},
\end{eqnarray*}
with $\Delta_{\sigma\mu\nu}=(-ie)\{\delta_{\mu\nu}(-k-p-q)_\sigma +
\delta_{\nu\sigma}(p-q+p-q-k)_\mu +\delta_{\sigma\mu} (-p+q+k+k)_\nu \}$.

On the other hand, Feynman rules yield, for Fig. 1(b),
\begin{eqnarray*}
{1\over (2\pi)^4} \int d^4q \cdot {\bar u}(p',s')\cdot &i \cdot {1\over 2 \sqrt 2}
{e\over sin \theta_W}\cdot i \gamma_\lambda (1+ \gamma_5)\nonumber\\
\cdot {1\over i} {m_2-i\gamma\cdot q\over {m_2^2+q^2-i\epsilon}}\cdot
&i\cdot i \eta (-)u_0 \cdot {1\over i} {m_1-i\gamma\cdot q\over
{m_1^2 + q^2-i\epsilon}} \nonumber\\
\cdot i\cdot {1\over 2 \sqrt 2}{e\over sin\theta_W}\cdot &i \gamma_{\lambda'}
(1+\gamma_5)\cdot \nonumber\\
\cdot {1\over i} {\delta_{\lambda\lambda'}\over {M_W^2+(p'-q)^2-i\epsilon}}
\cdot {1\over i} {m_\mu - i\gamma\cdot p'\over {m_\mu^2+ p^{\prime 2}-i\epsilon}}
\cdot &i (-i)e \cdot \gamma\cdot {\epsilon(k)\over \sqrt {2k_0}}. u(p,s),
\end{eqnarray*}
and a similar result for Fig. 1(c).

The four-dimensional integrations can be carried out, via the dimensional
integration formulae (e.g. Ch. 10, Wu/Hwang \cite{Books}), especially
if we drop the small masses compared to the W-boson mass $M_W$ in the
denominator. In this way, we obtain
\begin{eqnarray*}
i T_a={G_F\over \sqrt 2} &\cdot \eta u_0
\cdot (m_1 + m_2)\cdot (-2i){e\over (4\pi)^2}\nonumber\\
&\cdot {\bar u}(p',s') {\gamma\cdot \epsilon\over \sqrt {2k_0}}
(1+\gamma_5) u(p,s).
\end{eqnarray*}

It is interesting to note that the wave-function renormalization,
as shown by Figs. 1(b) and 1(c), yields
\begin{eqnarray*}
i T_{b+c}= {G_F\over \sqrt 2} &\cdot \eta u_0 (m_1 + m_2)
\cdot (+2i){e\over (4\pi)^2}
\cdot \{{p'^2\over m_\mu^2 + p'^2} + {p^2\over m_e^2 + p^2}\}\nonumber\\ &\cdot
{\bar u}(p',s') {\gamma \cdot \epsilon \over \sqrt{2k_0}} (1+\gamma_5) u(p,s),
\end{eqnarray*}
noting that $p^2=-m_\mu^2$ and $p'^2=-m_e^2$ would make the contribution
from Figs. 1(b) and 1(c) to be of the opposite sign to that from Fig. 1(a).

It is interesting to note that the leading terms all cancel, a result of
gauge invariance. We have computed some next-order terms but a complete
result seems to be rather difficult to obtain.

In a normal treatment, one ignores the wave-function renormalization
diagrams 1(b) and 1(c) in the treatment of the decays $\mu \to e + \gamma$,
$\mu \to 3e$, and $\mu+ A \to e+ A^*$.

Comparing this to the dominant mode $\mu \to e {\bar \nu}_e \nu_\mu$
\cite{Books}, we could obtain the branching ratio.
Even though the decay rate for $\mu \to e+ \gamma$ would be of the order
$O(m_{neutrino}^4/M_W^4)$, which is extremely small. Note that
the cancelation does not exist for $\mu \to e + e^+ + e^-$, nor for
the conversion process $\mu^- + p \to e^- + p$. So, the rates would be
expected to be much larger.

The off-diagonal mass matrix would be modified by the self-energy
diagram since the neutrinos form a triplet under $SU_f(3)$. It is
presumed that these self-energy diagrams, after the ultraviolet
divergences get subtracted, lead to masses of the right order. If
the off-diagonal mass matrix is diagonalized alone, the three roots
would be two negative and one positive, adding up to zero. This seems
like one ordering in the masses of neutrinos - one up and two downs.

Besides the golden decay $\mu \to e+ \gamma$ (much too small) and
neutrino oscillations (already observed), violation of the
$\tau-\mu-e$ universality is also expected and might be observed. As
the baryon asymmetry is sometime attributed to the lepton-antilepton
asymmetry, the current scenario for neutrino oscillations \cite{PDG}
seems to be inadequate. If we take the hints from neutrinos rather
seriously, there are so much to discover, even though the minimal
Standard Model for the ordinary-matter world remains to be pretty 
much intact.

\medskip

\section{The Questions}

Let us come back to look at the statement of the extended Standard Model.
We choose the basic units at first and then the gauge group. The Higgs
mechanism would be in the last step.

If that is the case, we have some difficulty in writing down the
left-right model \cite{Salam}. why? If we need to assign a certain
left-handed or right-handed spinor into two basic units
simultaneously, then the kinetic term appears twice - our language does
not go; we believe that a lagrangian should have only one kinetic term.

So, our first question would be: Could the above rationale rule out the
right-handed sector, since the simultaneous presence of the left-handed
and right-handed basic units as $SU(2)$ doublets are excluded?
Experimentally, we should check this point. As long as we could argue,
we note that, as long as the left-handed and right-handed components are
split in the basic units, parity has to be violated, either V-A or V+A.

In a slightly different context \cite{Hwang3}, It was proposed that we
could work with two working rules: "Dirac similarity principle", based on
eighty years of experience, and "minimum Higgs hypothesis", from the last
forty years of experience. Using these two working rules, the extended
model mentioned above becomes rather unique - so, it is so much easier
to check it against the experiments.

The Model stated in the paper is yet to be completed, in view of the
"minimum Higgs hypothesis". The Higgs mechanism in the previous
$SU_f(3)$ family gauge theory is complete since the theory is treated
{\it alone}. With $SU_f(3)$ and $SU_L(2)$ (3, 2) Higgs multiplet
mentioned above plus one (3, 1) Higgs triplet, is it sufficient to do
the Higgs-mechanism job - no "unwanted" massless particles? we would
like to list this "mathematical" question as the second important
question.

We would be curious about how the dark-matter world looks like, though
it is difficult to verify experimentally. The first question would be: The
dark-matter world, 25 \% of the current Universe (in comparison, only 5 \%
in the ordinary matter), would clusterize to form the dark-matter
galaxies, maybe even before the ordinary-matter galaxies. The dark-matter
galaxies would then play the hosts of (visible) ordinary-matter galaxies,
like our own galaxy, the Milky Way. Note that a dark-matter galaxy is
by our definition a galaxy that does not possess any ordinary strong and
electromagnetic interactions (with our visible ordinary-matter world).
This fundamental question deserves some thoughts, for the structural
formation of our Universe.

Of course, we should remind ourselves that, in our ordinary-matter world,
those quarks can aggregate in no time, to hadrons, including nuclei, and
the electrons serve to neutralize the charges also
in no time. Then atoms, molecules, complex molecules, and so on. These serve as
the seeds for the clusters, and then stars, and then galaxies, maybe in a time span
of $1\, Gyr$ (i.e., the age of our young Universe). The aggregation caused by
strong and electromagnetic forces is fast enough to help giving rise to galaxies
in a time span of $1\, Gyr$. On the other hand, the seeded clusterings might
proceed with abundance of extra-heavy dark-matter particles such as familons
and family Higgs, all greater than a few $TeV$ and with relatively long
lifetimes (owing to very limited decay channels). So, further simulations on
galactic formation and evolution may yield clues on our problem.

So, we could put forward the third important question of this paper:
What are the details of the dark-matter world?

Finally, coming back to the fronts of particle physics, neutrinos,
especially the right-handed neutrinos, might couple to the dark-matter
particles. Any further investigation along this direction would be of
utmost importance. It may shed light on the nature of the dark-matter world.

\medskip

\parindent=0pt
{\bf Notes added when we decided to publish the paper:}
\parindent=12pt

The minimum Higgs picture is as follows: the Standard-Model Higgs
doublet $\Phi(1,2)$, the purely family Higgs triplet $\Phi(3,1)$, 
and the mixed family Higgs $\Phi(3,2)$. The triplet $\Phi(3,1)$ and 
the neutral part $\Phi^0(3,2)$ undergoes the spontaneous symmetry 
breaking as in \cite{Family}. In the U-gauge, the neutral part 
$\Phi^0{3,2}$ gets projected out and the mass term becomes negative 
while there remains to have no SSB in the charged part $\Phi^+(3,2)$.
For details, please consult \cite{Hwang12}.     

\bigskip

Over the years, Pauchy Hwang would like to thank Jen-Chieh Peng and
Tony Zee for numerous interactions, those, plus a lot of (unspoken)
personal thoughts, lead to this paper. This work is supported in
part by National Science Council project (NSC99-2112-M-002-009-MY3).

\end{document}